\documentclass[prb,twocolumn,preprintnumbers,superscriptaddress]{revtex4}

\usepackage{graphicx}
\usepackage{dcolumn}
\usepackage{amsmath}
\usepackage{color}

\begin{document}

\title{Electronic correlations and spin-charge-density stripes in double-layer La$_3$Ni$_2$O$_7$}

\author{I. V. Leonov}
\affiliation{M. N. Mikheev Institute of Metal Physics, Russian Academy of Sciences, 620108 Yekaterinburg, Russia}
\affiliation{Institute of Physics and Technology, Ural Federal University, 620002 Yekaterinburg, Russia}
\affiliation{Center for Photonics and 2D Materials, Moscow Institute of Physics and Technology, 141700 Dolgoprudny, Russia}

\begin{abstract}
Using \emph{ab initio} band structure and DFT+dynamical mean-field theory methods we examine the effects of 
electron-electron interactions on the electronic structure, magnetic state, and structural phase stability of the recently discovered double-layer perovskite superconductor La$_3$Ni$_2$O$_7$ (LNO) under pressure.
Our results show the emergence of a double spin-charge-density stripe 
state characterized by a wave vector ${\bf q}=(\frac{1}{4},\frac{1}{4})$ arrangement of the nominally high-spin Ni$^{2+}_\mathrm{A}$ and low-spin Ni$^{3+}_\mathrm{B}$ ions (diagonal hole stripes oriented at $45^\circ$ to the Ni-O bond) which form zigzag ferromagnetic chains alternating in the $ab$ plane. 
The phase transition is accompanied by cooperative breathing-mode distortions of the lattice structure and leads to a reconstruction of the low-energy electronic structure and magnetic properties of LNO. We obtain a narrow-gap correlated insulator with a band gap value of $\sim$0.2 eV characterized by strong localization of the Ni $3d$ states and significant spin-orbital polarizations of the charge deficient Ni$^{3+}_\mathrm{B}$ ions.
Our results suggest the importance of double exchange to determine the magnetic properties of LNO, similarly to that in charge-ordered manganites.
We propose that spin and charge stripe fluctuations play an important role to tune superconductivity in LNO under pressure.
 
\end{abstract}

\maketitle

\section{Introduction}

The discovery of superconductivity in the rare-earth infinite-layer nickelates (Nd,Sr)NiO$_2$ in 2019 by D. Li \emph{et al.} \cite{Li_2019} has stimulated intensive experimental and theoretical efforts to understand the unusual properties of this novel class of superconducting materials \cite{Hepting_2020,Osada_2021,Lee_2023}. It has been demonstrated that superconductivity sets below $\sim$15 K in the high-quality single crystalline thin films $R$NiO$_2$ ($R$ is a rare-earth element, Sr, and Ca). 
While being structurally and electronically similar to the hole-doped cuprate superconductors \cite{Azuma_1992,Peng_2017}, 
the Ni $3d$ states (with the $x^2-y^2$ orbital states dominating near the Fermi level) experience strong hybridization with the $R$ $5d$ bands leading to self-doping \cite{Lee_2004,Kitatani_2020,Chen_2022a,Nomura_2022,Botana_2022}. As a result, the Fermi surface (FS) of $R$NiO$_2$ has multiorbital character distinct from that in cuprates. Moreover, low-energy excitations in the infinite-layer nickelates can be characterized as Mott-Hubbard-like, with strong orbital-dependent correlations, in contrast to a charge-transfer regime in the high-$T_c$ cuprates \cite{Lechermann_2020,Werner_2020,Karp_2020a,Lechermann_2020b,Wang_2020,Leonov_2020,
Nomura_2020,Talantsev_2020,Leonov_2021,Wan_2021,Karp_2022,Kreisel_2022,Pascut_2023,Talantsev_2023,Worm_2024,Cataldo_2024}. It is interesting that despite experimental evidence of antiferromagnetic correlations, the parent phases of these materials do not exhibit long-range magnetic ordering down to $\sim$2~K \cite{Lu_2021,Lin_2021,Cui_2021,Zhou_2022,Lin_2022,Fowlie_2022,Ortiz_2022,Krieger_2024}. In addition, experiments suggest the emergence of charge density wave in the uncapped hole-doped $R$NiO$_2$ films \cite{Rossi_2021,Tam_2021,Krieger_2021}, implying the complex interplay between the spin and charge degrees of freedoms in these materials  \cite{Slobodchikov_2022,Chen_2023a,Shen_2023,Onari_2023,Peng_2023,Stepanov_2024,Alvarez_2024,
RZhang_2024}.

In comparison to cuprates, the infinite-layer nickelates are characterized by a significantly reduced critical temperature, $T_c \sim 15$ K, with a modest increase up to $\sim$31 K upon optimizing their crystalline quality, compositions, epitaxial lattice strain, and pressure \cite{Wang_2022,Ren_2023}.
In this respect, the recent discovery of superconductivity (SC) in the bulk double-layer perovskite nickelate La$_3$Ni$_2$O$_7$ (LNO) with the Ruddlesden-Popper (RP) structure La$_{n+1}$Ni$_n$O$_{3n+1}$ ($n=2$) with a transition temperature comparable to that in cuprates, $T_c \sim 80$ K, under pressure above $\sim$15 GPa \cite{Sun_2023,Hou_2023,Wang_2024a,YZhang_2024,Liu_2023a} has ignited great interest to this novel class of superconducting materials \cite{Dong_2024,Yang_2024a,Liu_2024,Xie_2024a,Chen_2024a,Kakoi_2024,Agrestini_2024,Dan_2024,Khasanov_2024,Xie_2024b,
Shilenko_2023,Lechermann_2023,Christiansson_2023,Yang_2023,Zhang_2023a,Shen_2023b,
Sakakibara_2023a,Cao_2024,Chen_2023b,Wu_2024,Ryee_2024,Heier_2024,Fan_2024,
Lechermann_2024,Liao_2023,Liu_2023b,Tian_2024a,Lu_2024,Craco_2024}.
It is also worth noting that SC with $T_c \sim 20$ K has recently been observed in the bulk trilayer nickelate La$_4$Ni$_3$O$_{10}$ (RP phase with $n=3$) under pressure above 15 GPa \cite{Sakakibara_2024,Zhu_2024,Li_2024a,Li_2024b}, implying that the multilayer mixed-valent RP nickelates host pressure-driven SC. 

Both the double-layer and trilayer LNO systems belong to the RP series La$_{n+1}$Ni$_n$O$_{3n+1}$ with $n = 2$ and 3, respectively, with two-dimensional perovskite-like slabs of corner-sharing NiO$_6$ octahedra interleaved with La-O units \cite{Ling_2000}. 
While structurally related to the infinite-layer nickelates (in the former all the apical oxygens are removed applying soft chemistry
topotactic reduction of the parent perovskite phase) \cite{Li_2019,Hepting_2020,Osada_2021,Lee_2023}, the electronic state of these materials is sufficiently different. First, both materials are intrinsically mixed 
valent. Thus, for the bilayer material Ni ions adopt a nominal Ni$^{2.5+}$ $3d^{7.5}$ electronic configuration, characterized by an equal amount of Ni$^{2+}$ and 3+ ions in the unit cell (for its stoichiometric composition). For the trilayer material it is Ni$^{2.67+}$ with two Ni$^{3+}$ and one Ni$^{2+}$ ions per formula unit. Moreover, in these materials both the Ni $x^2-y^2$ and $3z^2-r^2$ orbitals contribute to the Fermi surface.

Interestingly, single-crystal x-ray diffraction, 
transport, muon spin relaxation ($\mu^+$ SR), and NMR measurements show the emergence of spin and charge density wave state in LNO upon cooling below 150 K (at ambient pressure) \cite{Sakakibara_2024,Zhu_2024,Li_2024a,
Li_2024b,Wu_2001,Chen_2024a,Kakoi_2024,Agrestini_2024,
Xie_2024b,Dan_2024,Khasanov_2024}, implying the complex interaction between spin and charge degrees of freedoms in LNO. Moreover, this result corroborates with recent resonant inelastic x-ray scattering and inelastic neutron diffraction measurements which show the presence of the in-plane magnetic correlations associated with the spin-charge density wave stripe state \cite{Agrestini_2024,Xie_2024b}. It is remarkable that superconductivity in LNO with $n=2$ and 3 is found to appear near a pressure-driven structural transition to the high-symmetry (orthorhombic) phase, which is characterized by the absence of tilting of oxygen octahedra \cite{Sun_2023,Liu_2023a}. 
It seems that SC in LNO is associated with suppression of a long-range spin and charge density wave state under pressure. 

At low pressures SC competes with a weakly insulating behavior \cite{
Sun_2023,Wang_2024a,Hou_2023,YZhang_2024,Liu_2023a,Wu_2001}. Moreover, above $T_c$ SC turns to a bad metal phase, implying the importance of strong correlations effects. In agreement with this, the many-body DFT+dynamical mean-field theory (DFT+DMFT) \cite{Georges_1996,Kotliar_2006} and $GW$+DMFT \cite{Biermann_2003,Tomczak_2017} electronic structure calculations show the importance of orbital-dependent correlations, such as orbital-selective quasiparticle mass renormalizations $m^*/m$ and incoherence of the spectral weights of the Ni $3d$ states in LNO \cite{Shilenko_2023,Lechermann_2023,Christiansson_2023,Ryee_2024,
Liao_2023,Heier_2024,Lechermann_2024,Craco_2024,Cao_2024,
Tian_2024a,Wang_2024,Leonov_2024}. Moreover, these studies propose the crucial importance of the effects of Fermi surface nesting and lattice to explain the low-energy structure of these materials (e.g., the calculated FSs exhibit multiple in-plane nesting effects). This behavior suggests the possible emergence of competing long-range spin and charge ordering effects (e.g., the formation of competing spin and charge density wave stripe states) at low pressure \cite{Shilenko_2023,Lechermann_2023,Leonov_2024,Lechermann_2024}. This implies that spin and charge stripe fluctuations can be crucial for the understanding of the low-energy physics of superconducting nickelates.

While the double-layer and trilayer LNO have recently been studied using various computational techniques \cite{Shilenko_2023,Lechermann_2023,Christiansson_2023,Yang_2023,Zhang_2023a,Shen_2023b,
Sakakibara_2023a,Cao_2024,Chen_2023b,Wu_2024,Ryee_2024,Heier_2024,Fan_2024,
Lechermann_2024,Liao_2023,Liu_2023b,Tian_2024a,Lu_2024,
Craco_2024,Li_2017,Wang_2024,LaBollita_2024a,Leonov_2024,ZhangLin_2024b,Huang_2024,
Chen_2024b,Yang_2024b}, the microscopic understanding of their unusual properties still poses a great challenge. In fact, the details of their electronic structure, magnetic state, and superconducting properties, as well as the key properties, such as microscopic origins of SC are still poorly understood. We address these remarkable topics in our present study.

In our paper, we discuss the electronic and magnetic properties of the normal state of the bilayer nickelate superconductor La$_3$Ni$_2$O$_7$. Using the DFT+U \cite{Anisimov_1991,Liechtenstein_1995,Dudarev_1998} and DFT+DMFT methods we study the electronic structure, magnetic state, orbital-dependent correlation effects, and structural 
phase stability of this system. In particular, we focus on the possible formation of the in-plane long-range spin and charge density wave stripe states, and 
on associated with this behavior reconstructions of the electronic structure and magnetic properties of LNO. Our results suggest that both strong correlations 
and lattice effects (e.g., cooperative breathing-mode lattice distortions caused by the formation of spin-charge-density wave stripes, SCDW) play a key role for understanding the low-energy electronic structure of the double-layer LNO.
We propose that the emergence of competing spin and charge stripe states and, hence, of spin-charge stripe fluctuations, is crucial for understanding of superconductivity in this material.

\section{RESULTS AND DISCUSSION}

\subsection{Computational details}

In this work, we discuss the effects of electronic correlations on the electronic and magnetic properties of the bilayer LNO. As a starting point, we compute its electronic structure, magnetic state, and structural phase stability using the DFT+U method \cite{Anisimov_1991,Liechtenstein_1995,Dudarev_1998}. In DFT we use generalized gradient approximation with the Perdew-Burke-Ernzerhof exchange-correlation functional and ultrasoft pseudopotentials as implemented in the Quantum ESPRESSO software package \cite{Giannozzi_2009, Giannozzi_2017, DalCorso_2014}. In our calculations, to account for correlation effects in the partially occupied Ni $3d$ shell we apply an effective Hubbard $U$ value ($U_\mathrm{eff} \equiv U-J$) ranging from the noninteracting DFT ($U=0$ eV) to 5 eV. The latter is a typical upper-bound effective Hubbard $U$ value used in the electronic structure calculations of nickelates. The DFT+U calculations are performed for the $2\sqrt{2} a \times \sqrt{2} b \times c$ $Cmmm$ supercell (containing 48 atoms) of the $Fmmm$ crystal structure of LNO. The lattice parameters $a=5.289$ \AA, $b=5.218$ \AA, and $c=19.734$ \AA\ were taken from experiment \cite{Sun_2023}. Using 
the DFT+U method we perform structural optimization of the internal atomic positions. Moreover, we perform full structural optimization of LNO and evaluate the crystal structure parameters of LNO at ambient pressure (using DFT+U).

We compute the electronic state and magnetic structure of LNO (with otimized crystal structure) using the DFT+U approach formulated within the basis of the localized Wannier functions constructed for the partially filled Ni $3d$ and O $2p$ valence states (constructed using the atomic-centred symmetry-constrained Wannier functions technique \cite{Anisimov_2005,Marzari_2012}). In these calculations we use the average Hubbard parameters $U = 4$ and 6 eV and Hund's exchange 
coupling $J = 0.95$ eV to treat static correlation effects in the partially filled Ni $3d$ shell \cite{Liechtenstein_1995}. 

Furthermore, we explore the effects of dynamical correlations of the Ni $3d$ electrons on the electronic structure, magnetic state, and exchange couplings of LNO. To this end, for the Wannier Ni $3d$ and O $2p$ states we construct the effective tight-binding Hubbard model \cite{Anisimov_2005,Marzari_2012} supplemented with the on-site Coulomb interaction for the partially filled Ni $3d$ electrons (parametrized with the average Hubbard $U$ and Hund's exchange $J$). This allows us to take into account strong correlations of the Ni $3d$ electrons, complicated by a charge transfer between the partially occupied Ni $3d$ and O $2p$ states.
We solve this effective many-body model using the DMFT method \cite{Georges_1996,Kotliar_2006} implemented with the continuous-time hybridization expansion (segment) quantum Monte-Carlo algorithm (with the Hubbard $U=6$ eV and Hund's exchange $J=0.95$ eV for the Ni $3d$ orbitals) \cite{Gull_2011}. 

\subsection{Electronic structure}

\begin{figure}[tbp!]
\centerline{\includegraphics[width=0.5\textwidth,clip=true]{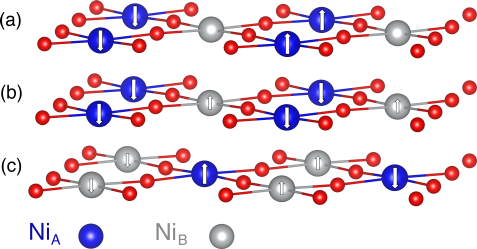}}
\caption{A scheme of the spin and charge ordering stripe patterns inside the Ni-O plane with ``Ni$^{2+}$'' ions shown in blue (Ni$_\mathrm{A}$) and charge deficient ``Ni$^{3+}$'' ions depicted in grey (Ni$_\mathrm{B}$) as calculated using the DFT+U method. In (a) the single spin-charge stripe arrangement is shown; (b) charge stripe; (c) double spin-charge stripe. The calculated Ni $3d$ spin moments are depicted by arrows. In (a) the single spin-charge stripe pattern is characterized by the alternating high-spin down/ non-magnetic/ high-spin up/ non-magnetic arrangement of the Ni$_\mathrm{A}$ and Ni$_\mathrm{B}$ ions. In (b) and (c) it consists of the alternating high-spin Ni$^{2+}_\mathrm{A}$ and low-spin Ni$^{3+}_\mathrm{B}$ states depicted by large and small arrows, respectively. We use the VESTA software to visualize the crystal structures. \cite{Vesta}
}
\label{Fig_1}
\end{figure}

In our DFT+U calculations we compute the electronic and magnetic properties for various types of long-range magnetic orderings of LNO. We examine the phase stability of the ferromagnetic (FM), $C$-type and $G$-type antiferromagnetic (AF) phases of LNO. In addition, 
we explore the possible emergence of the long-range spin and charge density wave stripe phases as depicted in Fig.~\ref{Fig_1}. We consider the single spin-charge stripe, charge stripe, and double spin-charge stripe phases which are depicted in Fig.~\ref{Fig_1}. These striped phases are characterized by the in-plane long-range ordering of the Ni ions moments in the basal $ab$ in-plane of the NiO$_6$ bilayer with a propagating wave vector ${\bf q}=(\frac{1}{4},\frac{1}{4})$, oriented at $45^\circ$ to the Ni-O bond. It corresponds to a ${\bf q}=(\frac{1}{2}, 0)$ arrangement in orthorhombic notation. In Fig.~\ref{Fig_1} the out-of-plane spin moments arrangement of the Ni ions (along the $c$-axis in the NiO$_6$ bilayer) is AF due to the large overlap between the two Ni $3z^2-r^2$ orbitals. 

It is also worth noting that similar in-plane density wave modulations (although with different in-plane periodicity) were previously discussed in the context of mixed-valence layered and hole-doped infinite-layer nickelates \cite{Lee_1997,Yoshizawa_2000,Zhang_2019,Zhang_2020}. Moreover, recent resonant inelastic x-ray scattering and inelastic neutron diffraction measurements of magnetic excitation in LNO exhibit the possible formation of the in-plane spin- and charge density wave states in LNO below $\sim$150 K (at low pressure) \cite{Agrestini_2024,Xie_2024b}. 
In agreement with this, previous analysis of magnetic correlations within DFT+DMFT propose the competing spin-charge-density waves states in the hole-doped infinite-layer and double-layer nickelates \cite{Slobodchikov_2022,Shilenko_2023,Lechermann_2023,Lechermann_2024}. In particular, for the former system the DFT+U and DFT+DMFT calculations show the emergence of the spin-charge-density and bond disproportionation stripe phases (oriented at $45^\circ$ to the Ni-O bond with an antiphase domain boundary of hole stripes) accompanied by in-plane breathing-mode distortions of the crystal structure \cite{Slobodchikov_2022}. 

\begin{figure}[tbp!]
\centerline{\includegraphics[width=0.5\textwidth,clip=true]{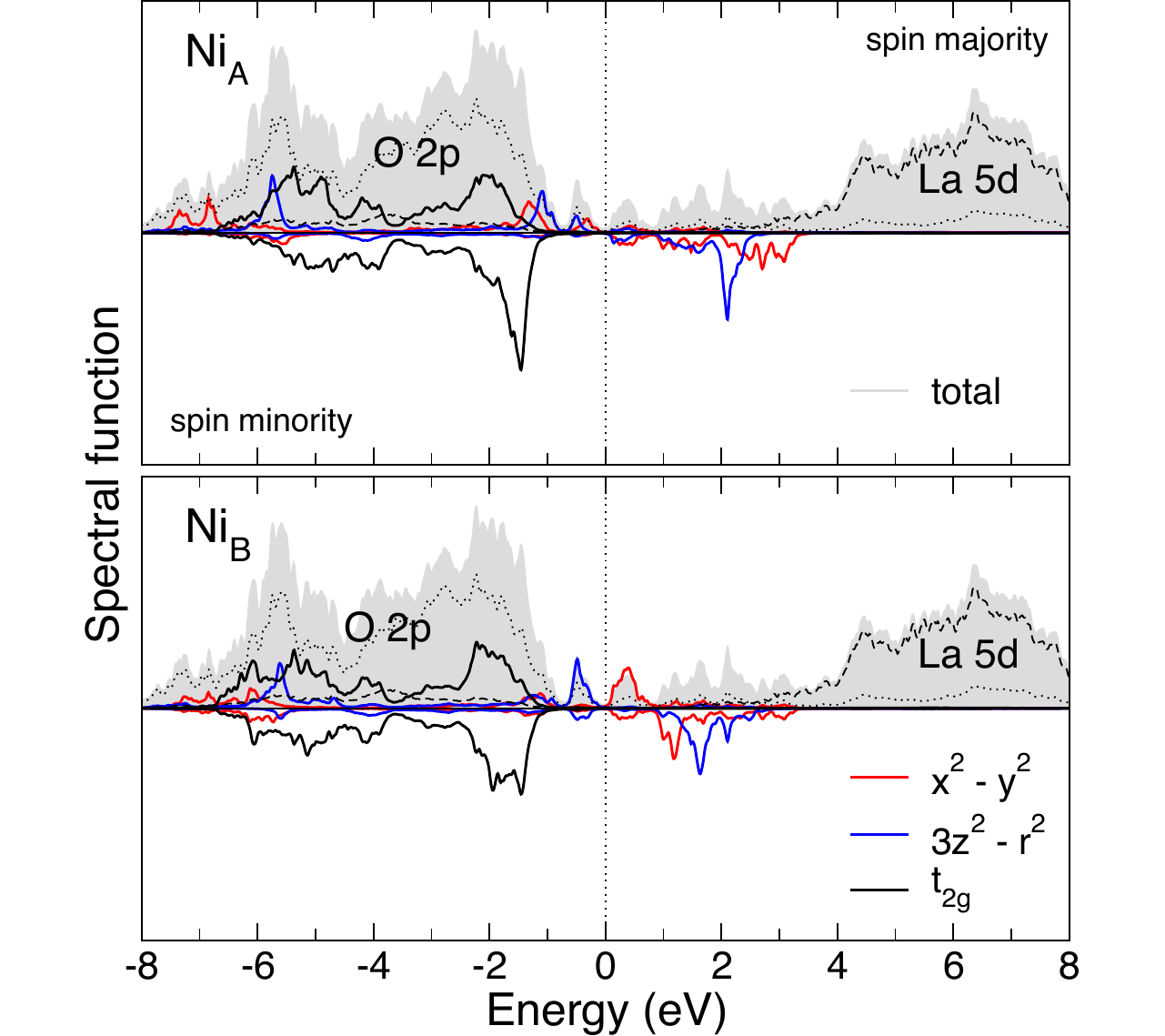}}
\caption{Orbital-dependent spectral functions of the double SCDW phase of LNO obtained by DFT+U with the Hubbard $U=4$ eV and Hund's exchange $J=0.95$ eV. The partial Ni $t_{2g}$, $x^2 - y^2$, and $3z^2 - r^2$ orbital contributions are shown (per orbital). The partial Ni $t_{2g}$, $x^2 - y^2$, and $3z^2 - r^2$ orbital states are magnified by a factor 4 for better readability.
}
\label{Fig_2}
\end{figure}

We perform structural optimization and compute the electronic structure and magnetic properties of LNO using the DFT+U method with different Hubbard $U_\mathrm{eff}$ values ($U_\mathrm{eff}=0$, 3, and 5 eV). Our DFT+U total energy calculations show that the double SCDW state (depicted in Fig.~\ref{Fig_1}c) has the lowest total energy among the considered magnetic configurations, implying its thermodynamic phase stability. We therefore focus on the properties of this phase.

Structural optimization leads to cooperative breathing-mode distortions of the lattice structure, implying a phase transformation of the ``undistorted'' $Fmmm$ phase of LNO with a symmetry reduction. We observe a remarkable difference in the average Ni-O bond length from that in the ``undistorted'' crystal structure. Note that in the $Fmmm$ phase Ni sites are structurally equivalent and the average Ni-O bond length is $\sim$1.87 \AA. In contrast to this, the calculated crystal structure is characterized by the emergence of two structurally distinct Ni sites, with ``expanded'' (Ni$_\mathrm{A}$) and ``contracted'' (Ni$_\mathrm{B}$) oxygen octahedra, with the average Ni-O distances 1.928 and 1.904 \AA, respectively (as obtained for $U_\mathrm{eff}=3$ eV). Moreover, we note a sufficient elongation of the ``contracted'' Ni$_\mathrm{B}$O$_6$ octahedra along the apical $c$-axis, implying the possible orbital polarization (orbital ordering) of the $3d$ electrons.

In Fig.~\ref{Fig_2} we display our results for the orbital-dependent spectral functions of the double SCDW phase of LNO obtained using the DFT+U method with the Hubbard $U=4$ eV and Hund's exchange $J=0.95$ eV. 
We find a narrow-gap magnetically ordered insulator with an energy gap of $\sim$0.3 eV, consistent with a weakly insulating behavior obtained for LNO in resistivity measurements. The calculated total Ni $3d$ Wannier occupancies at the Ni A and B sites are about 8.35 and 8.23, respectively. Our results imply the emergence of charge disproportionation between the Ni A and B $3d$ states, of 0.12. 
In agreement with this the calculated within DFT+U Ni $3d$ spin moments are 1.32 and 0.8 $\mu_B$ for the Ni$_\mathrm{A}$ and Ni$_\mathrm{B}$ sites, respectively. 
This suggests that the Ni$_\mathrm{A}$ ions adopt the high-spin Ni$^{2+}$ state (with a $3d^8$ electronic configuration), while the Ni$_\mathrm{B}$ ions are being depopulated close to the Ni$^{3+}$ ($3d^7$) low-spin state. 

An analysis of the majority spin Ni $e_g$ Wannier orbital occupations confirms substantial charge disproportionation in the double SCDW phase. In fact, the majority spin $x^2-y^2$ and $3z^2-r^2$ orbital occupations are about 0.89 and 0.94 for the Ni$_\mathrm{A}$ and 0.56 and 0.96 for the Ni$_\mathrm{B}$ ions, respectively (as obtained by DFT+U for $U=4$ eV and $J=0.95$ eV). At the same time, the corresponding Ni $e_g$ Wannier charges difference (including the spin majority and minority states) is only about 0.12. Moreover, for the Ni$_\mathrm{B}$ ions we observe a sufficient spin-orbital polarization of the majority spin $e_g$ orbitals, $p=|(n_{x^2-y^2} - n_{z^2})/(n_{x^2-y^2} + n_{z^2})| \sim 0.26$, with a predominant occupation of the majority spin $3z^2-r^2$ states ($n_{x^2-y^2}$ and $n_{z^2}$ are the corresponding orbital occupations). Our result for the orbital polarization including the spin majority and minority Ni$_\mathrm{B}$ $e_g$ states is sufficiently smaller, about 0.16. In contrast to this, the $x^2-y^2$ and $3z^2-r^2$ majority spin orbital occupations of the Ni$^{2+}$ A ions are nearly same (with $p\sim 0.03$). 

A large spin-orbital polarization of the Ni$_\mathrm{B}$ $e_g$ states is consistent with an analysis of the local distortions of the NiO$_6$ oxygen octahedra. We find that for Ni$_\mathrm{B}$ sites the average out-of-plane Ni-O bond length (2.042 \AA) is considerably, by $\sim$10\% larger than the average Ni-O distance in the basal $ab$ plane (1.835 \AA). 
Interestingly, it is even larger than the (average) out-of-plane Ni-O bond length for the Ni$^{2+}$ ions, 2.024 \AA. For comparison, the average Ni-O bond length for the Ni$_\mathrm{A}$ ions is about 1.928 \AA.
We also note the importance of the Ni $3d$ and O $2p$ charge transfer, which leads to a reduction of orbital polarization of the Ni$^{3+}_\mathrm{B}$ ion (which has a fully occupied spin-majority $3z^2-r^2$ orbital state). In fact, it leads to suppression of the orbital polarization to $p \sim 0.26$, which is sufficiently smaller of its ionic value. This result is in agreement with strong hybridization between the Ni $3d$ and O $2p$ states.

Our results show that the Wannier Ni $3d$ spin moments and charge disproportionation between the Ni A and B sites depend sensitively upon the Hubbard $U$ value. In particular, for $U=6$ eV and $J=0.95$ eV the spin 
moments are 1.52 and 0.95 $\mu_\mathrm{B}$ for the Ni A and Ni B ions, respectively (for the double SCDW phase). The calculated total Ni $3d$ Wannier occupancies for the Ni A and B ions 
are about 8.3 and 8.17, respectively. We note that the double SCDW phase is found to be thermodynamically stable for all Hubbard $U$ values. 
Moreover, the spin-charge stripe phase with the alternating in-plane arrangement of the high-spin Ni$^{2+}_\mathrm{A}$ $S=1$ and low-spin Ni$_\mathrm{B}$ $S=\frac{1}{2}$ ions depicted in Fig.~\ref{Fig_1}b (with the Ni $3d$ spin moments of 1.2 and 0.4 $\mu_\mathrm{B}$, respectively) is found to be less stable, with a total energy difference of 70 meV/f.u. The single spin-charge stripe arrangement (Fig.~\ref{Fig_1}a) is by about 75 meV/f.u. higher in total energy. 
The FM, $C$-type and $G$-type AF configurations are by more than 100 meV/f.u. higher in the total energy than the double SCDW phase (all for $U=4$ eV and $J=0.95$ eV) .

\subsection{Site- and orbital-dependent correlations}

Next, we explore the effects of dynamical correlations of the Ni $3d$ electrons on the electronic structure and magnetic state of LNO using the DFT+DMFT method. In our study we focus on the the double SCDW phase of LNO. In the DFT+DMFT 
calculations we use the distorted crystal structure of LNO obtained upon full structural optimization within DFT+U (with the effective Hubbard $U$ value of 5 eV). 
We construct the effective low-energy Hamiltonian for the Wannier Ni $3d$ and O $2p$ states using a basis set of atomic-centered Wannier functions within the energy window spanned by these bands \cite{Anisimov_2005}.
In order to treat correlation effects in the partially occupied Ni $3d$ shell we apply the Hubbard $U = 6$ and Hund's exchange $J = 0.95$ eV for the Ni $3d$ orbitals.
In DMFT the realistic many-body problem is solved using the continuous-time hybridization expansion (segment) quantum Monte Carlo algorithm (CT-QMC) \cite{Gull_2011}. In CT-QMC we solve two distinct impurity problems for the structurally different Ni sites. To treat AF ordering of the Ni ions we impose symmetry constraint to the spin majority and spin minority self-energies at different Ni A and B sites. In DFT+DMFT we utilize the around mean-field double-counting correction evaluated from the self-consistently determined local occupations. In order to compute the {\bf k}-resolved spectra we perform analytic continuation of the self-energy results using Pad\'e approximants. Our DFT+DMFT calculations are performed for the long-range AF state of LNO at a temperature $T = 116$ K (for the double SCDW phase depicted in Fig.~\ref{Fig_1}c).

\begin{figure}[tbp!]
\centerline{\includegraphics[width=0.5\textwidth,clip=true]{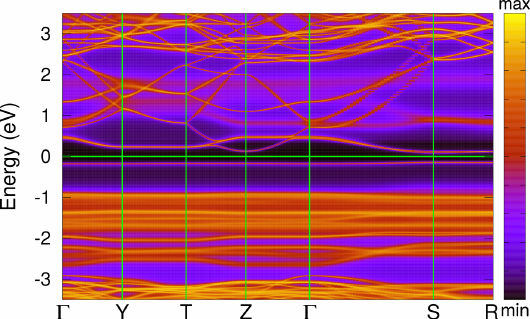}}
\caption{{\bf k}-resolved spectral function of the double SCDW phase of LNO calculated using DFT+DMFT with $U=6$ eV and $J=0.95$ eV at $T=116$~K. 
}
\label{Fig_3}
\end{figure}

\begin{figure}[tbp!]
\centerline{\includegraphics[width=0.5\textwidth,clip=true]{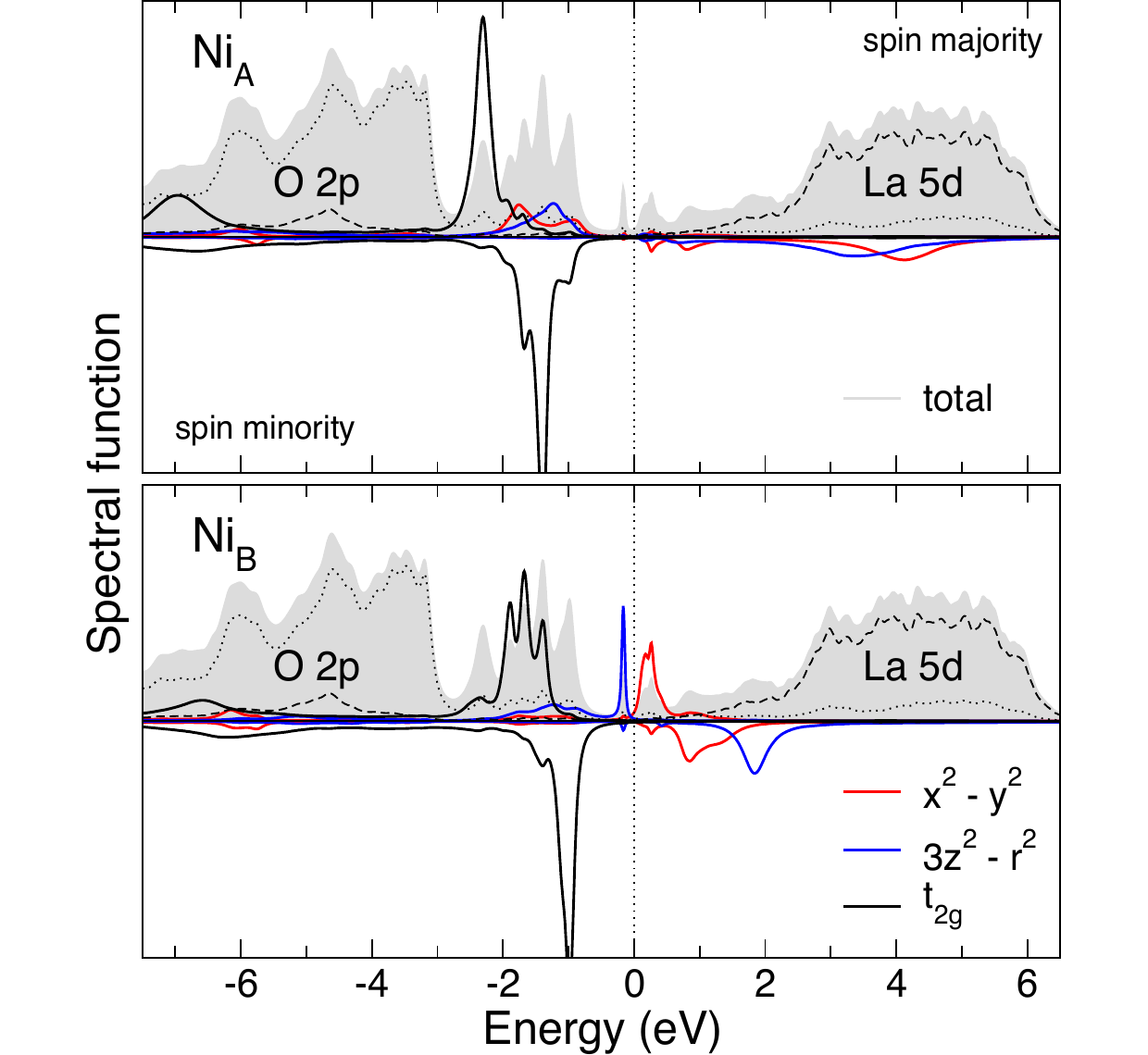}}
\caption{Orbitally resolved spectral functions obtained by DFT+DMFT with $U=6$ eV and $J=0.95$ eV for the double SCDW phase of LNO ($T=116$ K). 
}
\label{Fig_4}
\end{figure}

In Fig.~\ref{Fig_3} we display the {\bf k}-resolved spectral function of the double SCDW phase of LNO obtained using the DFT+DMFT approach (with $U = 6$ and $J = 0.95$ eV). Our results for the orbital-resolved spectral functions of LNO are summarized in Fig.~\ref{Fig_4}. 
In comparison to the previously published DFT+DMFT results for the $Fmmm$ phase of PM LNO (see, e.g., Refs.~\onlinecite{Shilenko_2023,Lechermann_2023}), we obtain a large redistribution of the Ni $3d$ and O $2p$ spectral weights near the Fermi level. In qualitative agreement with our DFT+U calculations, we obtained a narrow-gap insulator with an energy gap of $\sim$0.2 eV.
The Ni$_\mathrm{A}$ $e_g$ orbital states are nearly fully spin-polarized and show a large splitting of about 4-5 eV between the (fully occupied) majority spin states located at about -1.5 eV below the Fermi level and (empty) minority spin states at about 2-4 eV above $E_F$. The Ni$_\mathrm{A}$ $t_{2g}$ orbital states are fully occupied and are located at about -2.2 eV for the majority and at -1.5 eV for the minority spin states, respectively.
An energy gap opens between the top of the valence band predominantly formed by the majority spin Ni$_\mathrm{B}$ $3z^2-r^2$ states (strongly hybridizing with the O $2p$ orbitals) and the empty majority spin $x^2-y^2$ orbitals. The minority spin Ni$_\mathrm{B}$ $e_g$ orbital states are empty and appear at about 1-2 eV above the Fermi level.
The O $2p$ states show a dominant contribution between -8 and -2.5 eV below $E_F$. The La $5d$ states are empty and appear above 2 eV.

It is important to note that the occupied majority spin Ni$_\mathrm{B}$ $3z^2-r^2$ orbital states form a nearly flat dispersionless band just below the Fermi level, resembling a localized spin-polaron state. The bottom of the conduction band is primarily of the majority spin Ni$_\mathrm{B}$ $x^2-y^2$ orbital character, with a narrow bandwidth of about 0.6 eV.
Overall, this implies the crucial importance of correlation effects upon doping of the double SCDW phase of LNO. The latter would lead either to a hole doping of the flat majority spin Ni $3z^2-r^2$ band or to an electron doping (e.g., due to introducing oxygen vacancies) of the Ni $x^2-y^2$ orbital states.

The spin-polarized DFT+DMFT calculations yield large ordered magnetic moments (on-site magnetization of the Ni $3d$ orbitals) at the Ni A and B sites. In fact, the calculated static on-site magnetic moment at the Ni$_\mathrm{A}$ ions is about 1.69 $\mu_\mathrm{B}$, implying that Ni$_\mathrm{A}$ ions are in the high-spin magnetic state $S = 1$.
Note that at the Ni$_\mathrm{B}$ site the static magnetic moment is sufficiently smaller, $\sim$0.95 $\mu_\mathrm{B}$ ($S = \frac{1}{2}$). This suggests that the Ni A and B ions have different spin and charge states, in accordance with the DFT+U results. Indeed, the majority spin Ni $x^2-y^2$ and $3z^2-r^2$ Wannier orbital occupations are 0.93 and 0.97 for the Ni$_\mathrm{A}$ and are 0.36 and 0.98 for the Ni$_\mathrm{B}$ ions, respectively (as obtained using the spin-polarized DMFT calculations). Our results therefore suggest that Ni$_\mathrm{B}$ ions have a nominal low-spin Ni$^{3+}$ $3d^7$ electronic configuration with nearly empty the majority spin Ni $x^2-y^2$ orbital, in agreement with a spin-charge density wave phase obtained within DFT+U.

\begin{figure}[tbp!]
\centerline{\includegraphics[width=0.5\textwidth,clip=true]{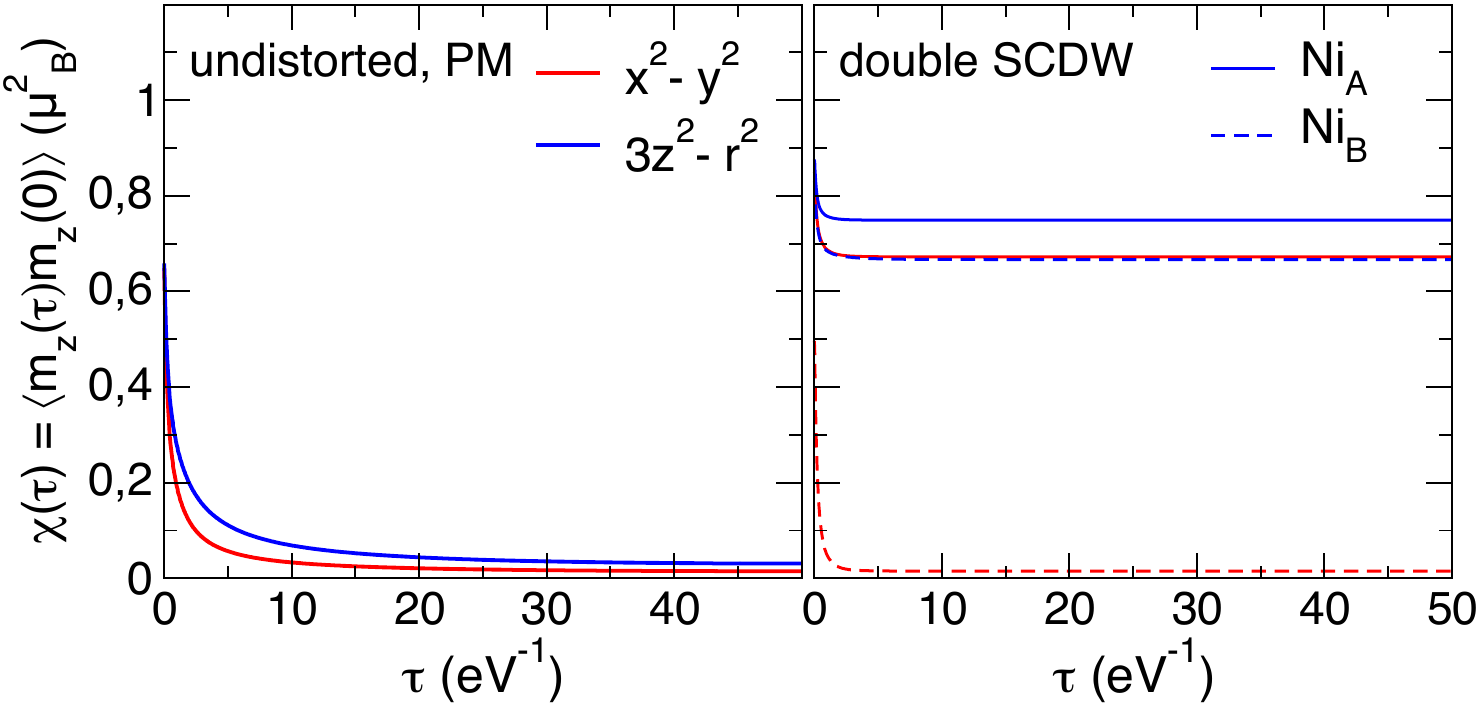}}
\caption{
Orbital- and site-dependent local spin-spin correlation functions $\chi(\tau) = \langle
\hat{m}_z(\tau) \hat{m}_z(0)\rangle$ as a function of the imaginary time $\tau$ for the Ni $e_g$ orbitals calculated by DFT+DMFT with $U=6$ eV and $J=0.95$ eV at $T = 116$ K. Left: Our results for the undistorted $Fmmm$ crystal structure of LNO in the PM state. Right: $\chi(\tau)$ for the long-range magnetically ordered SCDW phase. 
}
\label{Fig_5}
\end{figure}

Moreover, this result agrees with sufficiently different instantaneous local magnetic moments $\sqrt{\hat{m}^2_z} \simeq 1.78$ $\mu_\mathrm{B}$ and 1.25 $\mu_\mathrm{B}$ obtained by DFT+DMFT for the Ni$_\mathrm{A}$ and Ni$_\mathrm{B}$ ions, respectively. The corresponding fluctuating magnetic moments evaluated as $M_\mathrm{loc} \equiv [k_\mathrm{B}T \int \chi(\tau) d\tau]^{1/2}$ (where $\chi(\tau) \equiv \langle \hat{m}_z(\tau)\hat{m}_z(0) \rangle $ 
is the local spin-spin correlation function evaluated within DMFT on the imaginary time $\tau$ domain) are 1.69 $\mu_\mathrm{B}$ and 0.94 $\mu_\mathrm{B}$. This suggests strong localization of the Ni $3d$ states in the double SCDW phase. For comparison in Fig.~\ref{Fig_5} we display our DFT+DMFT results for the undistorted (correlated metallic) $Fmmm$ phase as obtained by DFT+DMFT (with full charge self-consistency) at $T=116$ K with the same Coulomb $U=6$ eV and Hund's exchange $J=0.95$ eV \cite{Shilenko_2023}. For the $Fmmm$ phase we obtain a sufficiently smaller fluctuating magnetic moment $M_\mathrm{loc} \simeq 0.41$ $\mu_\mathrm{B}$, with the instantaneous moment of $\sim$1.32 $\mu_\mathrm{B}$.(Note that Ni ions are equivalent in the undistorted $Fmmm$ phase. This implies the crucial importance of cooperative distortions of the lattice and long-range magnetic ordering in LNO.
In fact, for the double SCDW phase $\chi(\tau)$ for the Ni$_\mathrm{A}$ $x^2-y^2$ and $3z^2-r^2$ orbitals is seen to be large and nearly independent of $\tau$, implying strong localization of the Ni $e_g$ orbitals. It also suggests a higher degree of correlation effects (and, hence, localization) for the Ni$_\mathrm{A}$ $3z^2-r^2$ orbital, which is presumably associated with by $\sim$20\% more narrow DFT bandwidth of the Ni$_\mathrm{A}$ $3z^2-r^2$ (about 3 eV) than that for the $x^2-y^2$ states ($\sim$3.6 eV). For the Ni$_\mathrm{B}$ ions $\chi(\tau)$ exhibits strong localization of the occupied $3z^2-r^2$ orbital, while the $x^2-y^2$ states are nearly empty.

An analysis of the weights of different atomic configurations of the Ni $3d$ electrons (in DMFT the Ni $3d$ electrons are seen fluctuating between various atomic configurations) gives 0.04, 0.77, and 0.18 for the atomic Ni $3d^7$, $d^8$, and
$d^9$ multiples for the Ni$_\mathrm{A}$ and 0.32, 0.57, and 0.11 for the Ni$_\mathrm{B}$ ions, respectively.
Our results for the spin-state configurations 
suggest strong interplay of the Ni $S = 0$, $\frac{1}{2}$ , and 1 states in the electronic structure of the double SCDW phase of LNO. Thus, for the Ni$_\mathrm{A}$ ion the corresponding spin-state weights are 0.04, 0.22, and 0.73. For the Ni$_\mathrm{B}$ ions these are 0.27, 0.43, and 0.29. This result is consistent with the high-spin Ni$_\mathrm{A}^{2+}$ $S=1$ and low-spin Ni$_\mathrm{B}^{3+}$ $S=\frac{1}{2}$ states, as it was established above.

Most importantly, our DFT+DMFT calculations for the double SCDW phase give finite Ni $3d$ Wannier static spin magnetic moments at both the Ni A and B sites, $\sim$1.69 $\mu_\mathrm{B}$ and 0.95 $\mu_\mathrm{B}$, respectively. This suggests the emergence of a \emph{concomitant} spin- and charge density-wave state formed at the Ni$_\mathrm{A}$ and Ni$_\mathrm{B}$ ions. In fact, in the $ab$ plane spin moments at the neighboring Ni$^{2+}_\mathrm{A}$ and Ni$^{3+}_\mathrm{B}$ ions form zigzag FM chains, which are aligned AF in the $ab$-plane oriented at $45^\circ$ to the Ni-O bonds (see Fig.~\ref{Fig_1}). The double SCDW phase can be characterized by a ${\bf q}=(\frac{1}{4},\frac{1}{4})$ spin-charge density wave modulation [${\bf q}=(\frac{1}{2},0)$ in orthorhombic notation]. As a result, this behavior resembles the unique 
electronic state of the charge-ordered manganites \cite{Mori_1998,Radaelli_1999,Loudon_2005} and suggests the importance of double exchange mechanism to determine the magnetic properties 
of LNO \cite{Slobodchikov_2022}. It is interesting that a similar zigzag spin- and charge stripe state (denoted as a bond-disproportionation state) was previously discussed in the context of the hole-doped infinite-layer nickelates, $R$NiO$_2$ \cite{Slobodchikov_2022}, implying an intrinsic instability of the layered mixed-valent nickelates to a spin-charge-density wave ordering.

The obtained double SCDW state is consistent with our microscopic 
analysis of the Heisenberg exchange couplings calculated using the magnetic force theorem within DFT+DMFT (within the Ni $3d$ and O $2p$ Wannier basis set) \cite{Liechtenstein_1987,Kvashnin_2015,Korotin_2015}. Our DFT+DMFT calculations give a large AF out-of-plane exchange coupling between the Ni ions within the NiO$_6$ bilayer (i.e., within a structural Ni-Ni dimer), of about 16.1 and 49.8~meV for the Ni A and B ions, respectively. 
Note that we adopt the following notation for the Heisenberg
model, $H = - \sum_{i>j} J_{ij} {e}_i{e}_j$, where $e_i$ are the unit vectors.
In agreement with the Kugel-Khomskii theory \cite{Kugel_1973,Kugel_1982}, this behavior is associated with a predominant occupation of the Ni $3z^2-r^2$ orbitals for the Ni$_\mathrm{A}$ and Ni$_\mathrm{B}$ ions forming a spin-singlet-like state (for the Ni A and B ions the majority spin $3z^2-r^2$ orbitals are fully occupied). 
For comparison, our DFT+U calculations with $U=4$ eV and $J=0.95$ eV of the same exchange couplings yield 36.9 and 97.1~meV, respectively.
At the same time, the in-plane nearest-neighbor exchange couplings between the Ni$_\mathrm{A}$ and Ni$_\mathrm{B}$ ions are found to be sufficiently weaker, of about 8.8~meV. Moreover, the appearance of the zigzag FM chains of the neighboring Ni$_\mathrm{A}$ and Ni$_\mathrm{B}$ sites (belonging to the same zigzag FM chain) consistent with the in-plane FM coupling between the Ni$_\mathrm{A}^{2+}$ (with a fully occupied majority spin $x^2-y^2$ orbital) and the Ni$_\mathrm{B}^{3+}$ ions (with a nearly empty majority spin $x^2-y^2$ state), mediated by charge transfer effects.
In agreement with this, the in-plane nearest-neighbor exchange couplings between the Ni$_\mathrm{A}$ and Ni$_\mathrm{B}$ ions belonging to the neighboring zigzag Ni$_\mathrm{A}$-Ni$_\mathrm{B}$ chains are AF, of about 8.6~meV.

Overall, our results suggest the stability of the spin-charge density wave stripe state in LNO at low temperature. It is interesting to note 
that our DFT+DMFT calculations at $T=116$~K for the distorted crystal structure of LNO in the PM state yield a site-selective Mott phase with the Ni$_\mathrm{A}$ site being Mott localized while the Ni$_\mathrm{B}$ sites show correlated metallic behavior \cite{Park_2012,Greenberg_2018}.
This suggests the crucial importance of strong correlations and long-range magnetic ordering to explain an insulating behavior of the double SCDW phase of LNO. For the PM phase our DMFT calculations exhibit orbital-selective quasiparticle mass enhancement and incoherence (bad metal behavior) of the metallic Ni$_\mathrm{B}$ $e_g$ bands caused by correlation effects. An analysis suggests a Fermi-liquid-like behavior of the Ni$_\mathrm{B}$ $e_g$ self-energies with a remarkable orbital selectivity of the quasiparticle mass renormalizations, with $m^*/m \simeq 2.2$ and 12 for the Ni$_\mathrm{B}$ $x^2-y^2$ and $3z^2-r^2$ bands, respectively. The latter are
evaluated as 
$m^*/m=[1-\partial Im \Sigma(i\omega_n) \partial i\omega_n]|_{i\omega_n \rightarrow 0}$ using Pad\'e approximants. 
The Ni$_\mathrm{B}$ $x^2-y^2$ and $3z^2-r^2$ 
orbitals show quasiparticle damping of $Im[\Sigma(i\omega)] \sim 0.08$ and 0.3 eV at the first Matsubara frequency, at $T = 116$ K. Thus, the Ni $3z^2-r^2$ states are seen to be more correlated and incoherent like than the planar Ni $x^2-y^2$ orbitals, consistent with a larger by about 20\% bandwidth of the Ni $x^2-y^2$ states (obtained by nonmagnetic DFT). This result also shows the proximity of the Ni$_\mathrm{B}$ $e_g$ states to the orbital-selective Mott state.

Most importantly, we note a remarkable reconstruction of the electronic states of LNO near the Fermi level caused by the interplay of correlations effects and long-range charge-density wave stripe state. 
Our DFT+DMFT calculations of the double SCDW phase of LNO (for the AF state) show a charge disproportionation between the Ni A and B $3d$ Wannier states accompanied by a significant orbital polarization of the Ni$_\mathrm{B}$ $e_g$ orbitals. Thus, the DFT+DMFT calculations of the double SCDW phase yield for 
the total Ni $3d$ Wannier occupations for the Ni A and B ions of 8.11 and 7.73, respectively, implying a robust charge disproportionation of about 0.38. In agreement with the DFT+U results the spin majority Ni$_\mathrm{B}$ $e_g$ orbitals show a significant orbital polarization, $p \simeq 0.46$. The total orbital polarization of the Ni$_\mathrm{B}$ $e_g$ state is about 0.3. For the PM LNO (with the SCDW distorted crystal structure) we obtain a sufficiently smaller Ni $e_g$ orbital polarization 0.2. For the total Ni $3d$ Wannier charge disproportion between the Ni$_\mathrm{A}$ and Ni$_\mathrm{B}$ sites we obtain 0.31. 
 
We therefore conclude that the high-symmetry $Fmmm$ phase of LNO is unstable at low pressure and temperature conditions. It leads to a transition to the insulating state which is accompanied by breathing-mode distortions of the lattice and long-range spin-charge density wave stripe ordering. Upon pressure tuning, a phase transition into a metallic state of LNO sets in, associated with suppression of the double SCDW phase. It turns out that superconductivity in LNO appears near to this phase transformation, which is suggestive of the importance of spin and charge stripe fluctuations to explain superconductivity in this materials.
In agreement with this, our spin-polarized DFT+DMFT calculations (at $T=116$ K) suggest that a long-range magnetically ordered insulating state is fragile. Indeed, upon a decrease of the Hubbard $U$ to $4$ eV (with $J=0.95$ eV) we obtain a metallic solution. Moreover, the calculated static (ordered) magnetic moments are significantly reduced to about 1.2 and 0.1~$\mu_\mathrm{B}$ for the Ni A and B sites, respectively. 
In addition, our spin-polarized DFT+DMFT calculations with the fully localized double counting scheme with $U=6$ eV and $J=0.95$ eV, at $T=116$ K yield a site-selective Mott phase, with the Mott localized Ni$_\mathrm{A}$ and correlated metallic behavior of the Ni$_\mathrm{B}$ $3d$ electrons. It is accompanied by the suppression of static (long-range ordering) magnetic moments of the Ni$_\mathrm{B}$ ions. The calculated ordered magnetic moment at the Ni$_\mathrm{A}$ ions is about 1.1~$\mu_\mathrm{B}$.
We propose that in such calculations the N\'eel temperature associated with a long-range magnetic ordering of the Ni$_\mathrm{B}$ ions is below 116~K, implying the presence of a crossover between the single and double spin-charge stripe arrangements (see Figs.~\ref{Fig_1}a and \ref{Fig_1}c).

\section{Conclusion}

To conclude, our DFT+U and DFT+DMFT results reveal a remarkable instability of the high-symmetry $Fmmm$ phase of LNO towards the formation of a spin-charge-density wave stripe state. Our results yield a narrow-gap correlated insulator with an energy gap of $\sim$0.2 eV accompanied by cooperative breathing-mode distortions of the lattice structure, implying the complex interplay between electronic states and lattice degrees of freedom. 
This behavior results in a sufficient reconstruction of the low-energy electronic structure and magnetic properties of LNO. 
Our results show the emergence of a double spin-charge-density stripe 
state characterized by diagonal hole stripes oriented at $45^\circ$ to the Ni-O bond which form zigzag ferromagnetic chains alternating in the $ab$ plane, suggesting the importance of double exchange to determine the magnetic properties of LNO, similarly to that in charge-ordered manganites.
We note the crucial importance of correlation effects upon doping of the double SCDW phase of LNO. The latter would lead either to a hole doping of the flat majority spin Ni$_\mathrm{B}$ $3z^2-r^2$ band (spin-polaron-like behavior) or to an electron doping of the $x^2-y^2$ orbital states (e.g., due to introducing oxygen vacancies), with strong orbital dependence of quasiparticle mass renormalizations of these states.
Our analysis suggests strong localization of the Ni $3d$ orbitals in the insulating phase of LNO. 
%
%
We propose the importance of spin-charge-density wave stripe states to determine the electronic structure and magnetic state of LNO. It appears that spin and charge stripe fluctuations can play an important role for the understanding of superconductivity in nickelates. 
%

\emph{Note added.} We note that complementary independent DFT+U calculations have been reported in Refs.~\onlinecite{LaBollita_2024b,BZhang_2024,Ni_2024}. These results are in overall agreement with our conclusions.

\section{ACKNOWLEDGMENTS}

The DFT+DMFT calculations, theoretical analysis of the electronic structure
and magnetic properties were supported by the Russian Science Foundation (Project No. 24-12-00186). The DFT and DFT+U calculations were supported by the Ministry of Science and Higher Education of the Russian Federation, project No. 122021000038-7 (theme ``Quantum'').


\begin{thebibliography}{100}


\bibitem{Li_2019}
D. Li, K. Lee, B. Y. Wang, M. Osada, S. Crossley, H. R. Lee, Y. Cui, Y. Hikita, and H. Y. Hwang,
Nature \textbf{572}, 624 (2019). 

\bibitem{Hepting_2020}
M. Hepting, D. Li, C. Jia, H. Lu, E. Paris, Y. Tseng, X. Feng, M. Osada, E. Been, Y. Hikita \emph{et al.}, 
Nat. Mater. \textbf{19}, 381 (2020).

\bibitem{Osada_2021}
M. Osada, B. Y. Wang, B. H. Goodge, S. P. Harvey, K. Lee, D. Li, L. F. Kourkoutis, and H. Y. Hwang, 
Adv. Mater. \textbf{33}, 2104083 (2021).

\bibitem{Lee_2023}
K. Lee, B. Y. Wang, M. Osada, B. H. Goodge, T. C. Wang, Y. Lee, S. Harvey, W. J. Kim, Y. Yu, C. Murthy, S. Raghu, L. F. Kourkoutis, and H. Y. Hwang, 
Nature \textbf{619}, 288 (2023).


\bibitem{Azuma_1992}
M. Azuma, Z. Hiroi, M. Takano, Y. Bando, and Y. Takeda, 
Nature (London) \textbf{356}, 775 (1992).

\bibitem{Peng_2017}
Y. Y. Peng, G. Dellea, M. Minola, M. Conni, A. Amorese, D. Di Castro, G. M. De Luca, K. Kummer, M. Salluzzo, X. Sun, X. J. Zhou, G. Balestrino, M. Le Tacon, B. Keimer, L. Braicovich, N. B. Brookes, and G. Ghiringhelli, 
Nat. Phys. \textbf{13}, 1201 (2017).


\bibitem{Lee_2004}
K.-W. Lee and W. E. Pickett,
Phys. Rev. B \textbf{70}, 165109 (2004).


\bibitem{Kitatani_2020}
M. Kitatani, L. Si, O. Janson, R. Arita, Z. Zhong, and K. Held, 
npj Quantum Mater. \textbf{5}, 59 (2020).

\bibitem{Botana_2022}
A. S. Botana, K.-W. Lee, M. R. Norman, V. Pardo, and W. E. Pickett, 
Front. Phys. \textbf{9}, 813532 (2022).

\bibitem{Chen_2022a}
H. Chen, A. Hampel, J. Karp, F. Lechermann, and A. J. Millis, 
Front. Phys. \textbf{10}, 835942 (2022).

\bibitem{Nomura_2022}
Y. Nomura and R. Arita, 
Rep. Prog. Phys. \textbf{85}, 052501 (2022).



\bibitem{Lechermann_2020}
F. Lechermann,
Phys. Rev. X \textbf{10}, 041002 (2020).

\bibitem{Werner_2020}
P. Werner and S. Hoshino, 
Phys. Rev. B \textbf{101}, 041104(R) (2020).

\bibitem{Karp_2020a}
J. Karp, A. S. Botana, M. R. Norman, H. Park, M. Zingl, and A. Millis, 
Phys. Rev. X \textbf{10}, 021061 (2020).

\bibitem{Lechermann_2020b}
F. Lechermann, 
Phys. Rev. B \textbf{101}, 081110(R) (2020).

\bibitem{Wang_2020}
Y. Wang, C.-J. Kang, H. Miao, and G. Kotliar, 
Phys. Rev. B \textbf{102}, 161118(R) (2020).

\bibitem{Leonov_2020}
I. Leonov, S. L. Skornyakov, and S. Y. Savrasov, 
Phys. Rev. B \textbf{101}, 241108(R) (2020).

\bibitem{Nomura_2020}
Y. Nomura, T. Nomoto, M. Hirayama, and R. Arita, 
Phys. Rev. Research \textbf{2}, 043144 (2020).

\bibitem{Leonov_2021}
I. Leonov, 
J. Alloys Compd. \textbf{883}, 160888 (2021).

\bibitem{Wan_2021}
X. Wan, V. Ivanov, G. Resta, I. Leonov, and S. Y. Savrasov, 
Phys. Rev. B \textbf{103}, 075123 (2021).

\bibitem{Karp_2022}
J. Karp, A. Hampel, and A. J. Millis, 
Phys. Rev. B \textbf{105}, 205131 (2022).

\bibitem{Kreisel_2022}
A. Kreisel, B. M. Andersen, A. T. R\o{}mer, I. M. Eremin, and F. Lechermann,
Phys. Rev. Lett. \textbf{129}, 077002 (2022).

\bibitem{Pascut_2023}
G. L. Pascut, L. Cosovanu, K. Haule, and K. F. Quader, 
Commun. Phys. \textbf{6}, 45 (2023).


\bibitem{Worm_2024}
P. Worm, Q. Wang, M. Kitatani, I. Bia\l{}o, Q. Gao, X. Ren, J. Choi, D. Csontosov\'a, K.-J. Zhou, X. Zhou, Z. Zhu, L. Si, J. Chang, J. M. Tomczak, and K. Held, 
Phys. Rev. B \textbf{109}, 235126 (2024).

\bibitem{Cataldo_2024}
S. Di Cataldo, P. Worm, J. M. Tomczak, L. Si, and K. Held, 
Nat. Commun. \textbf{15}, 3952 (2024).

\bibitem{Talantsev_2020}
E. F. Talantsev, 
Results Phys. \textbf{17}, 103118 (2020).

\bibitem{Talantsev_2023}
E. F. Talantsev,
J. Appl. Phys. \textbf{134}, 113904 (2023).



\bibitem{Lu_2021}
H. Lu, M. Rossi, A. Nag, M. Osada, D. F. Li, K. Lee, B. Y. Wang, M. Garcia-Fernandez, S. Agrestini, Z. X. Shen, E. M. Been, B. Moritz, T. P. Devereaux, J. Zaanen, H. Y. Hwang, K.-J. Zhou, and W. S. Lee, 
Science \textbf{373}, 213 (2021).

\bibitem{Lin_2021}
J. Q. Lin, P. Villar Arribi, G. Fabbris, A. S. Botana, D. Meyers, H. Miao, Y. Shen, D. G. Mazzone, J. Feng, S. G. Chiuzba\' ian \emph{et al.}, 
Phys. Rev. Lett. \textbf{126}, 087001 (2021).

\bibitem{Cui_2021}
Y. Cui, C. Li, Q. Li, X. Zhu, Z. Hu, Y-F. Yang, J. Zhang, R. Yu, H.-H. Wen, and W. Yu, 
Chinese Phys. Lett. \textbf{38}, 067401 (2021).

\bibitem{Zhou_2022}
X. Zhou, X. Zhang, J. Yi, P. Qin, Z. Feng, P. Jiang, Z. Zhong, H. Yan, X. Wang, H. Chen \emph{et al.}, 
Adv. Mater. \textbf{34}, 2106117 (2022).

\bibitem{Lin_2022}
H. Lin, D. J. Gawryluk, Y. M. Klein, S. Huangfu, E. Pomjakushina, F. von Rohr, and A. Schilling, 
New J. Phys. \textbf{24}, 013022 (2022).

\bibitem{Fowlie_2022}
J. Fowlie, M. Hadjimichael, M. M. Martins, D. Li, M. Osada, B. Y. Wang, K. Lee, Y. Lee, Z. Salman, T. Prokscha, J.-M. Triscone, H. Y. Hwang, and A. Suter, 
Nat. Phys. \textbf{18}, 1043 (2022).

\bibitem{Ortiz_2022}
R. A. Ortiz, P. Puphal, M. Klett, F. Hotz, R. K. Kremer, H. Trepka, M. Hemmida, H.-A. Krug von Nidda, M. Isobe, R. Khasanov, H. Luetkens, P. Hansmann, B. Keimer, T. Sch\"afer, and M. Hepting,
Phys. Rev. Research \textbf{4}, 023093 (2022).

\bibitem{Krieger_2024}
G. Krieger, H. Sahib, F. Rosa, M. Rath, Y. Chen, A. Raji, P.V.B. Pinho, C. Lefevre, G. Ghiringhelli, A. Gloter, N. Viart, M. Salluzzo, and D. Preziosi,
arXiv:2403.16969 (2024).



\bibitem{Rossi_2021}
M. Rossi, M. Osada, J. Choi, S. Agrestini, D. Jost, Y. Lee, H. Lu, B. Y. Wang, K. Lee, A. Nag \emph{et al.}, 
Nat. Phys. \textbf{18}, 869 (2022).

\bibitem{Tam_2021}
C. C. Tam, J. Choi, X. Ding, S. Agrestini, A. Nag, B. Huang, H. Luo, M. Garc\' ia-Fern\' andez, L. Qiao, K.-J. Zhou, 
Nat. Mater. \textbf{21}, 1116 (2022).

\bibitem{Krieger_2021}
G. Krieger, L. Martinelli, S. Zeng, L. E. Chow, K. Kummer, R. Arpaia, M. M. Sala, N. B. Brookes, A. Ariando, N. Viart, M. Salluzzo, G. Ghiringhelli, D. Preziosi, 
Phys. Rev. Lett. \textbf{129}, 027002 (2022).



\bibitem{Slobodchikov_2022}
K. G. Slobodchikov and I. V. Leonov, 
Phys. Rev. B \textbf{106}, 165110 (2022).

\bibitem{Chen_2023a}
H. Chen, Y.-F. Yang, G.-M. Zhang, 
Nat. Commun. \textbf{14}, 5477 (2023).

\bibitem{Shen_2023}
Y. Shen, M. Qin, and G.-M. Zhang, 
Phys. Rev. B \textbf{107}, 165103 (2023).

\bibitem{Onari_2023}
S. Onari and H. Kontani,
Phys. Rev. B \textbf{108}, L241119 (2023).

\bibitem{Peng_2023}
C. Peng, H.-C. Jiang, B. Moritz, T. P. Devereaux, and C. Jia,
Phys. Rev. B \textbf{108}, 245115 (2023).

\bibitem{Stepanov_2024}
E. A. Stepanov, M. Vandelli, A. I. Lichtenstein, and F. Lechermann,
npj Comput. Mater \textbf{10}, 108 (2024).

\bibitem{Alvarez_2024}
\'A. A. C. \'Alvarez, L. Iglesias, S. Petit, W. Prellier, M. Bibes, and J. Varignon,
Phys. Rev. Materials \textbf{8}, 064801 (2024).

\bibitem{RZhang_2024}
R. Zhang, C. Lane, J. Nokelainen, B. Singh, B. Barbiellini, R. S. Markiewicz, A. Bansil, and J. Sun, 
Phys. Rev. Lett. \textbf{133}, 066401 (2024).


\bibitem{Wang_2022}
N. N. Wang, M. W. Yang, Z. Yang, K. Y. Chen, H. Zhang, Q. H. Zhang, Z. H. Zhu, Y. Uwatoko, L. Gu, X. L. Dong, K. J. Jin, J. P. Sun, and J.-G. Cheng, 
Nat. Commun. \textbf{13}, 4367 (2022).

\bibitem{Ren_2023}
X. Ren, J. Li, W.-C. Chen, Q. Gao, J. J. Sanchez, J. Hales, H. Luo, F. Rodolakis, J. L. McChesney, T. Xiang, J. Hu, R. Comin, Y. Wang, X. Zhou, and Z. Zhu, 
Commun. Phys. \textbf{6}, 341 (2023).


\bibitem{Sun_2023}
H. Sun, M. Huo, X. Hu, J. Li, Z. Liu, Y. Han, L. Tang, Z. Mao, P. Yang, B. Wang, J. Cheng, D.-X. Yao, G.-M. Zhang, and M. Wang,
Nature \textbf{621}, 493 (2023).

\bibitem{Liu_2023a}
Z. Liu, H. Sun, M. Huo, X. Ma, Y. Ji, E. Yi, L. Li, H. Liu, J. Yu, Z. Zhang, Z. Chen, F. Liang, H. Dong, H. Guo, D. Zhong, B. Shen, S. Li, and M. Wang,
Sci. China: Phys. Mech. Astron. \textbf{66}, 217411 (2023).

\bibitem{Hou_2023}
J. Hou, P. T. Yang, Z. Y. Liu, J. Y. Li, P. F. Shan, L. Ma, G. Wang, N. N. Wang, H. Z. Guo, J. P. Sun, Y. Uwatoko, M. Wang, G.-M. Zhang, B. S. Wang, and J.-G. Cheng,
Chin. Phys. Lett. \textbf{40}, 117302, (2023).

\bibitem{Wang_2024a}
G. Wang, N. N. Wang, X. L. Shen, J. Hou, L. Ma, L. F. Shi, Z. A. Ren, Y. D. Gu, H. M. Ma, P. T. Yang, Z.Y. Liu, H.Z. Guo, J.P. Sun, G.M. Zhang, S. Calder, J.-Q. Yan, B.S. Wang, Y. Uwatoko, and J.-G. Cheng,
Phys. Rev. X \textbf{14}, 011040 (2024).

\bibitem{YZhang_2024}
Y. Zhang, D. Su, Y. Huang, Z. Shan, H. Sun, M. Huo, K. Ye, J. Zhang, Z. Yang, Y. Xu, Y. Su, R. Li, M. Smidman, M. Wang, L. Jiao, and H. Yuan,
Nat. Phys. \textbf{20}, 1269 (2024).


\bibitem{Dong_2024}
Z. Dong, M. Huo, J. Li, J. Li, P. Li, H. Sun, L. Gu, Y. Lu, M. Wang, Y. Wang, and Z. Chen,
Nature \textbf{630}, 847 (2024).

\bibitem{Yang_2024a}
J. Yang, H. Sun, X. Hu, Y. Xie, T. Miao, H. Luo, H. Chen, B. Liang, W. Zhu, G. Qu, C.-Q. Chen, M. Huo, Y. Huang, S. Zhang, F. Zhang, F. Yang, Z. Wang, Q. Peng, H. Mao, G. Liu, Z. Xu, T. Qian, D.-X. Yao, M. Wang, L. Zhao, and X. J. Zhou,
Nat. Commun. \textbf{15}, 4373 (2024).

\bibitem{Liu_2024}
Z. Liu, M. Huo, J. Li, Q. Li, Y. Liu, Y. Dai, X. Zhou, J. Hao, Y. Lu, M. Wang, and H.-H. Wen,
Nat. Commun. \textbf{15}, 7570 (2024).

\bibitem{Xie_2024a}
T. Xie, M. Huo, X. Ni, F. Shen, X. Huang, H. Sun, H. C. Walker, D. Adroja, D. Yu, B. Shen, L. He, K. Cao, and M. Wang,
Sci. Bull. (2024), https://doi.org/10.1016/j.scib.2024.07.030


\bibitem{Chen_2024a}
K. Chen, X. Liu, J. Jiao, M. Zou, C. Jiang, X. Li, Y. Luo, Q. Wu, N. Zhang, Y. Guo, and L. Shu,
Phys. Rev. Lett. \textbf{132}, 256503 (2024).

\bibitem{Kakoi_2024}
M. Kakoi, T. Oi, Y. Ohshita, M. Yashima, K. Kuroki, T. Kato, H. Takahashi, S. Ishiwata, Y. Adachi, N. Hatada, T. Uda, and H. Mukuda
J. Phys. Soc. Jpn. \textbf{93}, 053702 (2024).



\bibitem{Xie_2024b}
T. Xie, M. Huo, X. Ni, F. Shen, X. Huang, H. Sun, H. C. Walker, D. Adroja, D. Yu, B. Shen, L. He, K. Cao, and M. Wang,
arXiv:2401.12635 (2024).

\bibitem{Agrestini_2024}
S. Agrestini, M. Garcia-Fernandez, X. Huang, H. Sun, D. Shen, M. Wang, J. Hu, Y. Lu, K.-J. Zhou, and D. Feng,
arXiv:2401.12657 (2024).

\bibitem{Dan_2024}
Z. Dan, Y. Zhou, M. Huo, Y. Wang, L. Nie, M. Wang, T. Wu, and X. Chen,
arXiv:2402.03952 (2024).



\bibitem{Khasanov_2024}
R. Khasanov, T. J. Hicken, D. J. Gawryluk, L. P. Sorel, S. B\"otzel, F. Lechermann, I. M. Eremin, H. Luetkens, and Z. Guguchia,
arXiv:2402.10485 (2024).




\bibitem{Shilenko_2023}
D. A. Shilenko and I. V. Leonov,
Phys. Rev. B \textbf{108}, 125105 (2023).

\bibitem{Lechermann_2023}
F. Lechermann, J. Gondolf, S. B\"otzel, and I. M. Eremin,
Phys. Rev. B \textbf{108}, L201121 (2023).

\bibitem{Christiansson_2023}
V. Christiansson, F. Petocchi, and P. Werner,
Phys. Rev. Lett \textbf{131}, 206501 (2023).

\bibitem{Ryee_2024}
S. Ryee, N. Witt, and T. O. Wehling,
Phys. Rev. Lett. \textbf{133}, 096002 (2024).

\bibitem{Liao_2023}
Z. Liao, L. Chen, G. Duan, Y. Wang, C. Liu, R. Yu, and Q. Si,
Phys. Rev. B \textbf{108}, 214522 (2023).

\bibitem{Heier_2024}
G. Heier, K. Park, and S. Y. Savrasov,
Phys. Rev. B \textbf{109}, 104508 (2024).

\bibitem{Lechermann_2024}
F. Lechermann, S. B\"otzel, and I. M. Eremin,
Phys. Rev. Materials \textbf{8}, 074802 (2024).

\bibitem{Craco_2024}
L. Craco and S. Leoni,
Phys. Rev. B \textbf{109}, 165116 (2024).

\bibitem{Cao_2024}
Y. Cao and Y.-F. Yang,
Phys. Rev. B \textbf{109}, L081105 (2024).

\bibitem{Tian_2024a}
Y.-H. Tian, Y. Chen, J.-M. Wang, R.-Q. He, and Z.-Y. Lu,
Phys. Rev. B \textbf{109}, 165154 (2024).


\bibitem{Yang_2023}
Q.-G. Yang, D. Wang, Q.-H. Wang,
Phys. Rev. B \textbf{108}, L140505 (2023).

\bibitem{Zhang_2023a}
Y. Zhang, L.-F. Lin, A. Moreo, and E. Dagotto,
Phys. Rev. B \textbf{108}, L180510 (2023).

\bibitem{Shen_2023b}
Y. Shen, M. Qin, and G.-M. Zhang,
Chin. Phys. Lett. \textbf{40}, 127401 (2023).

\bibitem{Sakakibara_2023a}
H. Sakakibara, N. Kitamine, M. Ochi, and K. Kuroki, Phys. Rev. Lett. \textbf{132}, 106002 (2024).

\bibitem{Chen_2023b}
X. Chen, P. Jiang, J. Li, Z. Zhong, and Y. Lu,
arXiv:2307.07154 (2023).

\bibitem{Wu_2024}
W. W\'u, Z. Luo, D.-X. Yao, and M. Wang,
Sci. China-Phys. Mech. Astron. \textbf{67}, 117402 (2024).

\bibitem{Fan_2024}
Z. Fan, J.-F. Zhang, B. Zhan, D. Lv, X.-Y. Jiang, B. Normand, and T. Xiang,
Phys. Rev. B \textbf{110} 024514 (2024).



\bibitem{Liu_2023b}
Y.-B. Liu, J.-W. Mei, F. Ye, W.-Q. Chen, and F. Yang,
Phys. Rev. Lett. \textbf{131}, 236002 (2023).



\bibitem{Lu_2024}
C. Lu, Z. Pan, F. Yang, and C. Wu,
Phys. Rev. Lett. \textbf{132}, 146002 (2024).


\bibitem{Sakakibara_2024}
H. Sakakibara, M. Ochi, H. Nagata, Y. Ueki, H. Sakurai, R. Matsumoto, K. Terashima, K. Hirose, H. Ohta, M. Kato, Y. Takano, and K. Kuroki,
Phys. Rev. B \textbf{109}, 144511 (2024).

\bibitem{Zhu_2024}
Y. Zhu, D. Peng, E. Zhang, B. Pan, X. Chen, L. Chen, H. Ren, F. Liu, Y. Hao, N. Li, Z. Xing, F. Lan, J. Han, J. Wang, D. Jia, H. Wo, Y. Gu, Y. Gu, L. Ji, W. Wang, H. Gou, Y. Shen, T. Ying, X. Chen, W. Yang, H. Cao, C. Zheng, Q. Zeng, J.-G. Guo, and J. Zhao
Nature \textbf{631}, 531 (2024).

\bibitem{Li_2024a}
J. Li, C.-Q. Chen, C. Huang, Y. Han, M. Huo, X. Huang, P. Ma, Z. Qiu, J. Chen, X. Hu, L. Chen, T. Xie, B. Shen, H. Sun, D.-X. Yao, and M. Wang,
Sci. China-Phys. Mech. Astron. \textbf{67}, 117403 (2024).

\bibitem{Li_2024b}
Q. Li, Y.-J. Zhang, Z.-N. Xiang, Y. Zhang, X. Zhu, H.-H. Wen,
Chinese Phys. Lett. \textbf{41}, 017401 (2024).

\bibitem{Wu_2001}
G. Wu, J. J. Neumeier, and M. F. Hundley,
Phys. Rev. B \textbf{63}, 245120 (2001).


\bibitem{Ling_2000}
C. D. Ling, D. N. Argyriou, G. Wu, and J. Neumeier,
J. Solid State Chem. \textbf{152}, 517 (2000).


\bibitem{Georges_1996}
A. Georges, G. Kotliar, W. Krauth, and M. J. Rozenberg,
Rev. Mod. Phys. \textbf{68}, 13 (1996).

\bibitem{Kotliar_2006}
G. Kotliar, S. Y. Savrasov, K. Haule, V. S. Oudovenko, O. Parcollet, and C. A. Marianetti,
Rev. Mod. Phys. \textbf{78}, 865 (2006).

\bibitem{Biermann_2003}
S. Biermann, F. Aryasetiawan, and A. Georges,
Phys. Rev. Lett. \textbf{90}, 086402 (2003).

\bibitem{Tomczak_2017}
J. M. Tomczak, P. Liu, A. Toschi, G. Kresse, and K. Held,
Eur. Phys. J. Special Topics \textbf{226}, 2565 (2017).


\bibitem{Wang_2024}
J.-X. Wang, Z. Ouyang, R.-Q. He, and Z.-Y. Lu,
Phys. Rev. B \textbf{109}, 165140 (2024).

\bibitem{Leonov_2024}
I. V. Leonov,
Phys. Rev. B \textbf{109}, 235123 (2024).






\bibitem{Li_2017}
H. Li, X. Zhou, T. Nummy, J. Zhang, V. Pardo, W. E. Pickett, J. F. Mitchell, and D. S. Dessau,
Nat. Commun. \textbf{8}, 704 (2017).


\bibitem{LaBollita_2024a}
H. LaBollita, J. Kapeghian, M. R. Norman, and A. S. Botana,
Phys. Rev. B \textbf{109} 195151 (2024).


\bibitem{ZhangLin_2024b}
Y. Zhang, L.-F. Lin, A. Moreo, T. A. Maier, and E. Dagotto,
Phys. Rev. Lett. \textbf{133}, 136001 (2024).

\bibitem{Huang_2024}
J. Huang and T. Zhou,
Phys. Rev. B \textbf{110}, L060506 (2024).

\bibitem{Chen_2024b}
C.-Q. Chen, Z. Luo, M. Wang, W. W\'u, and D.-X. Yao,
Phys. Rev. B \textbf{110}, 014503 (2024).

\bibitem{Yang_2024b}
Q.-G. Yang, K.-Y. Jiang, D. Wang, H.-Y. Lu, and Q.-H. Wang,
Phys. Rev. B \textbf{109}, L220506 (2024).





\bibitem{Anisimov_1991}
V. I. Anisimov, J. Zaanen, and O. K. Andersen,
Phys. Rev. B \textbf{44}, 943 (1991).

\bibitem{Liechtenstein_1995}
A. I. Liechtenstein, V. I. Anisimov, and J. Zaanen,
Phys. Rev. B \textbf{52}, R5467(R) (1995).

\bibitem{Dudarev_1998}
S. L. Dudarev, G. A. Botton, S. Y. Savrasov, C. J. Humphreys, and A. P. Sutton, 
Phys. Rev. B \textbf{57}, 1505 (1998).

\bibitem{Giannozzi_2009}
P. Giannozzi, S. Baroni, N. Bonini, M. Calandra, R. Car, C. Cavazzoni, D. Ceresoli, G. L. Chiarotti, M. Cococcioni, I. Dabo \emph{et al.}, 
J. Phys.: Condens. Matter \textbf{21}, 395502 (2009).

\bibitem{Giannozzi_2017}
P. Giannozzi, O. Andreussi, T. Brumme, O. Bunau, M. B. Nardelli, M. Calandra, R. Car, C. Cavazzoni, D. Ceresoli, M. Cococcioni \emph{et al.},
J. Phys.: Condens. Matter \textbf{29}, 465901 (2017).

\bibitem{DalCorso_2014} A. Dal Corso,
Comput. Mater. Sci. \textbf{95}, 337 (2014).


\bibitem{Anisimov_2005}
V. I. Anisimov, D. E. Kondakov, A. V. Kozhevnikov, I. A. Nekrasov, Z. V. Pchelkina, J. W. Allen, S.-K. Mo, H.-D. Kim, P. Metcalf, S. Suga, A. Sekiyama, G. Keller, I. Leonov, X. Ren, and D. Vollhardt,
Phys. Rev. B \textbf{71}, 125119 (2005).

\bibitem{Marzari_2012}
N. Marzari, A. A. Mostofi, J. R. Yates, I. Souza, and D. Vanderbilt,
Rev. Mod. Phys. \textbf{84}, 1419 (2012).


\bibitem{Gull_2011}
E. Gull, A. J. Millis, A. I. Lichtenstein, A. N. Rubtsov, M. Troyer, and P. Werner,
Rev. Mod. Phys. \textbf{83}, 349 (2011).

\bibitem{Vesta}
K. Momma and F. Izumi,
J. Appl. Crystallogr. \textbf{44}, 1272 (2011).


\bibitem{Lee_1997}
S.-H. Lee and S-W. Cheong,
Phys. Rev. Lett. \textbf{79}, 2514 (1997).

\bibitem{Yoshizawa_2000}
H. Yoshizawa, T. Kakeshita, R. Kajimoto, T. Tanabe, T. Katsufuji, and Y. Tokura,
Phys. Rev. B \textbf{61}, 854 (2000).


\bibitem{Zhang_2019}
J. Zhang, D. M. Pajerowski, A. S. Botana, H. Zheng, L. Harriger, J. Rodriguez-Rivera, J. P. C. Ruff, N. J. Schreiber, B. Wang, Y.-S. Chen, W. C. Chen, M. R. Norman, S. Rosenkranz, J. F. Mitchell, and D. Phelan,
Phys. Rev. Lett. \textbf{122}, 247201 (2019).

\bibitem{Zhang_2020}
J. Zhang, D. Phelan, A. S. Botana, Y.-S. Chen, H. Zheng, M. Krogstad, S. G. Wang, Y. Qiu, J. A. Rodriguez-Rivera, R. Osborn, S. Rosenkranz, M. R. Norman, and J. F. Mitchell,
Nat. Commun. \textbf{11}, 6003 (2020).







\bibitem{Mori_1998}
S. Mori, C. H. Chen, and S.-W. Cheong,
Nature (London) \textbf{392}, 473 (1998).

\bibitem{Radaelli_1999}
P. G. Radaelli, D. E. Cox, L. Capogna, S.-W. Cheong, and M. Marezio,
Phys. Rev. B \textbf{59}, 14440 (1999).

\bibitem{Loudon_2005}
J. C. Loudon, S. Cox, A. J. Williams, J. P. Attfield, P. B. Littlewood, P. A. Midgley, and N. D. Mathur,
Phys. Rev. Lett. \textbf{94}, 097202 (2005).


\bibitem{Liechtenstein_1987}
A. I. Liechtenstein, M. I. Katsnelson, V. P. Antropov, and V. A. Gubanov,
J. Magn. Magn. Mater. \textbf{67}, 65 (1987).

\bibitem{Kvashnin_2015}
Y. O. Kvashnin, O. Gr{\aa}n\"as, I. Di Marco, M. I. Katsnelson, A. I. Lichtenstein, and O. Eriksson,
Phys. Rev. B \textbf{91}, 125133 (2015).

\bibitem{Korotin_2015}
Dm. M. Korotin, V. V. Mazurenko, V. I. Anisimov, and S. V. Streltsov,
Phys. Rev. B \textbf{91}, 224405 (2015).


\bibitem{Kugel_1973}
K. I. Kugel and D. I. Khomskii,
Zh. Eksp. Teor. Fiz. \textbf{64},1429 (1973).

\bibitem{Kugel_1982}
K. I. Kugel and D. I. Khomskii, Sov. Phys. Usp. \textbf{25}, 231 (1982).



\bibitem{Park_2012}
H. Park, A. J. Millis, and C. A. Marianetti,
Phys. Rev. Lett. \textbf{109}, 156402 (2012).

\bibitem{Greenberg_2018}
E. Greenberg, I. Leonov, S. Layek, Z. Konopkova, M. P. Pasternak, L. Dubrovinsky, R. Jeanloz, I. A. Abrikosov, and G. Kh. Rozenberg,
Phys. Rev. X \textbf{8}, 031059 (2018).



\bibitem{LaBollita_2024b}
H. LaBollita, V. Pardo, M. R. Norman, A. S. Botana,
arXiv:2407.14409 (2024).

\bibitem{BZhang_2024}
B. Zhang, C. Xu, H. Xiang,
arXiv:2407.18473 (2024).

\bibitem{Ni_2024}
X.-S. Ni, Y. Ji, L. He, T. Xie, D.-X. Yao, M. Wang, and K. Cao,
arXiv:2407.19213 (2024).


\end{thebibliography}
\end{document}